\documentstyle[11pt,aaspp]{article}

\slugcomment{To appear in {\it The Astrophysical Journal} 10 April
1996} 
\begin{document}

\title{{\it HST} Observations of the Luminous IRAS Source
FSC10214+4724: \\ A Gravitationally Lensed Infrared
Quasar\altaffilmark{1}} \author{Peter R. Eisenhardt\altaffilmark{2},
Lee Armus\altaffilmark{3}, David W. Hogg\altaffilmark{3}, B. T.
Soifer\altaffilmark{3}, \\ G. Neugebauer\altaffilmark{3}, Michael W.
Werner\altaffilmark{2}}

\altaffiltext{1}{Based on observations made with the NASA/ESA Hubble
Space Telescope, obtained at the Space Telescope Science Institute,
which is operated by AURA under NASA contract NAS 5-26555.}
\altaffiltext{2}{MS 169-327, Jet Propulsion Laboratory, California
Institute of Technology, 4800 Oak Grove Drive, Pasadena, CA 91109}
\altaffiltext{3}{Division of Physics, Math, \& Astronomy, California
Institute of Technology, Pasadena, CA 91125}

\begin{abstract} With a redshift of 2.3, the IRAS source FSC 10214+4724
is apparently one of the most luminous objects known in the Universe.
We present an image of FSC10214+4724 at $0.8\mu$m obtained with the
Hubble Space Telescope WFPC2 Planetary Camera.  The source appears as
an unresolved ($< 0''\!.06$) arc $0''\!.7$ long, with significant
substructure along its length.  The center of curvature of the arc is
located near an elliptical galaxy $1''\!.18$ to the north.  An
unresolved component 100 times fainter than the arc is clearly detected
on the opposite side of this galaxy.  The most straightforward
interpretation is that FSC 10214+4724 is gravitationally lensed by the
foreground elliptical galaxy, with the faint component a counterimage
of the IRAS source.  The brightness of the arc in the {\it HST} image
is then magnified by $\sim100$ and the intrinsic source diameter is
$\sim0''\!.01 (80$ pc) at $0.25\mu$m rest wavelength.  The bolometric
luminosity is probably amplified by a smaller factor ($\sim 30$), due
to the larger extent expected for the source in the far-infrared.  A
detailed lensing model is presented which reproduces the observed
morphology and relative flux of the arc and counterimage, and correctly
predicts the position angle of the lensing galaxy.  The model also
predicts reasonable values for the velocity dispersion, mass, and
mass-to-light ratio of the lensing galaxy for a wide range of galaxy
redshifts.  A redshift for the lensing galaxy of $\sim0.9$ is
consistent with the measured surface brightness profile from the image,
as well as with the galaxy's spectral energy distribution.  The
background lensed source has an intrinsic luminosity $\sim 2 \times
10^{13} L_{\sun}$, and remains a highly luminous quasar with an
extremely large ratio of infrared to optical/ultraviolet luminosity.


\end{abstract}

\section{Introduction}

Ever since its identification with a redshift 2.286 optical emission
line source by Rowan-Robinson et al. (1991), leading to an inferred
bolometric luminosity $\sim5 \times 10^{14} L_{\sun}$, the IRAS source
FSC10214+4724 has been the subject of enormous attention.  Detections
of CO (Brown \& Vanden Bout 1991; Solomon, Downes \& Radford 1992;
Tsuboi \& Nakai 1992) and submillimeter continuum emission (Clements et
al. 1992, Downes et al. 1992) from the source confirmed the presence of
huge quantities of gas and dust.  With a vastly larger lookback time
and luminosity than any other known IRAS source, FSC10214+4724 appeared
to be either an extremely luminous dust embedded quasar, or a
representative of a new class of astronomical object, e.g. a primeval
galaxy.

However, while the redshift of the IRAS source is secure, its intrinsic
luminosity is less certain.  The fact that FSC10214+4724 lies at the
flux limit of the IRAS survey, combined with the presence of several
red companion objects within a few arcseconds, led Elston et al. (1994)
to suggest that the IRAS source might be gravitationally lensed by a
foreground group of galaxies.  Intriguingly, Matthews et al. (1994)
found arcs emerging from the source in a deconvolved $K$ band image
with $0''\!.6$ seeing taken with the Keck telescope.  Matthews et al.
considered the lensing hypothesis, but concluded it was unlikely
because the image morphology was not achromatic.  Broadhurst \& Leh\'ar
(1995) modelled the source as gravitationally lensed, finding support
for their model from a reanalysis of the Matthews et al. data.  Graham
\& Liu (1995) also argue for lensing, based on deconvolution of a more
recent (March 1995) Keck $K$ band image with $0''\!.4$ seeing.
Trentham (1995) argues on statistical grounds that magnification due to
lensing is likely to be less than a factor of ten, although larger
magnifications are reasonable for smaller far-IR source sizes than the
1 kpc Trentham assumed.

We present an image of FSC10214+4724 taken in December 1994 at 8000
\AA\ with the {\it HST} WFPC2 Planetary Camera with $0''\!.1$
resolution.  This image provides dramatic support for the lensing
hypothesis, implying a magnification in the {\it HST} data of $\sim
100$.  We use the image to derive a detailed model for the intrinsic
properties of the lensed source and the lensing galaxy.

For reference, at the FSC 10214+4724 redshift $z=2.286$, one
$0''\!.0455$ Planetary Camera pixel subtends $300(180) h^{-1}$pc for
$q_o = 0(0.5)$, while these values are $239(191) h^{-1}$pc for $z=0.9$,
where $h \equiv H_o/100$ km sec$^{-1}$Mpc$^{-1}$.  Where not otherwise
specified, we assume $H_o = 50$ km sec$^{-1}$Mpc$^{-1}$ and $q_o =
0.5$.

\section{Observations and Reduction}

Three frames, each 2200 seconds long, were obtained on consecutive
orbits with the WFPC2 F814W filter on the 10th and 11th of December
1994 (UT).  FSC 10214+4724 was positioned near the center of the
Planetary Camera, and each exposure was displaced from the other two by
an integer number (5 or 10) of PC pixels in both axes.  The Wide Field
Camera data are not considered here.

After standard processing provided by STScI, the multiple frames were
used to filter out cosmic rays and hot pixels.  Although these defects
are quite prominent and affect roughly 4\% of the pixels in each frame,
the main characteristics of the combined image discussed in
section~\ref{Morphology} are discernible in each frame even without
this filtering.

Cross-correlations were performed on pairs of frames to confirm that
the actual displacements between frames, as measured in pixels, were
integers to within 0.2 pixels.  The frames were then trimmed by the
appropriate number of rows and columns to coregister them, and the
STSDAS task CRREJ was used to average them together, iteratively
excluding pixels which deviated from the previous iteration's average
value by more than three sigma.  The minimum value at each pixel
location was used for the initial estimate of the average, and sigma
was the value expected from Poisson statistics and the gain and read
noise.  To remove multiple pixel cosmic ray events, a stricter limit of
1.5 sigma was applied to the four pixels adjacent to any pixel which
exceeded the three sigma criterion.  Finally a median filtering routine
was applied to identify and interpolate over a few dozen isolated
pixels which deviated sharply from their neighbors in the average
image, presumably because they were corrupted in all three frames.
None of these latter pixels fall within objects in the field, and only
a handful of the pixels in the components discussed below are based on
data from less than two frames.

\section{Results} \label{results}

The combined image of the full Planetary Camera field is shown in
Figure 1(a), while figures 1(b) and (c) show the FSC 10214+4724 region
in progressively greater detail.

A synthetic point spread function (PSF) derived from the ``Tiny Tim"
{\it HST} image modelling software package was used to deconvolve the
average image, because a good empirical point spread function was not
available (see section~\ref{Profiles}).  The synthetic PSF was
calculated for a source with the color of a K-star in F814W at the
location of FSC10214+4724 in the Planetary Camera field.   Figure 1(d)
shows the same region covered in figure 1(c) after a mild deconvolution
of the data (10 iterations of the STSDAS implementation of the
Lucy-Richardson algorithm) onto a grid subsampled four times more
finely than the original pixels.

\subsection{Morphology} \label{Morphology}

At the resolution of the Planetary Camera, an arc-like structure
dominates the morphology of the emission line source.  In the
terminology of Matthews et al. (1994), which is adopted here, the
arc-like structure is component 1 (see Fig. 1(b)).  The extent of this
arc is smaller than shown in Matthews et al., and there is a sharply
defined ridge of high surface brightness emission which is $0''\!.7$
long and essentially unresolved in the transverse direction.  Lower
surface brightness emission can be seen extending the arc $\sim0''\!.4$
to the west, and a similar amount (but at a considerably fainter level)
to the east-northeast.  There is also a hint of still fainter emission
extending a few tenths of an arcsecond due east ({\em not} along a
circular arc) from the eastern tip of the bright ridge.  Within the
bright ridge are at least two peaks separated by $0''\!.24$, with the
brighter peak towards the east.  The center of curvature of the arc was
fitted and found to be $\sim0''\!.12$ west-northwest of the center of
component 2 (which is $1''\!.18$ from the arc).  Component 2 has a
smooth light distribution which is resolved and slightly elongated (see
sections~\ref{Profiles} and \ref{Redshift}).  Directly opposite
component 2 from the arc is a faint but clearly visible source
(component 5 in figure 1(b)), $0''\!.43$ from the center of component
2.  Component 3 is resolved and has a feature which is suggestive of a
tidal arm leading back towards component 2. Component 4 appears to be a
highly inclined galaxy.

\subsection{Brightness Profiles} \label{Profiles}

In an attempt to quantify the radial extent of the arc, pixels from the
sector subtended by the brightest $0''\!.5$ of the arc at component 2
were sorted in order of radius from component 2.  To reduce the effect
of the tangential substructure along the arc, a running average of the
flux from 5 pixels in this radially sorted list was calculated.
Figure~2 plots this running average flux as a function of the average
radius of those pixels less the $1''\!.18$ distance of component 1 from
component 2.  For comparison, the (unaveraged) radial profiles are
plotted for stars A and H (see figure 1(a)), for components 2 and 5,
and for the synthetic PSF which was used in the deconvolution shown in
figure 1(d).

While the wings of the synthetic PSF fall inside those of the arc, the
{\em empirical} PSFs of stars A (outside its saturated core) and H
match the arc cross section reasonably well. It therefore appears
likely that the synthetic PSF underestimates the FWHM of the true PSF.
Based on the synthetic PSF we estimate an upper limit of $0''\!.06$
(500 pc) for the intrinsic FWHM of the arc in the radial direction.
Note that the effects of the running average, of any error in using
component~2 as the center of the arc, and of the smaller size of the
synthetic PSF all work in the direction of leading us to overestimate
this dimension.

The deconvolved image shown in Figure 1(d) also yields a $0''\!.06$
FWHM for the arc, but this holds true for star H after deconvolution as
well.  Because the individual frames are separated by integer numbers
of PC pixels, there is little leverage on finer scale structure.
Deconvolution does emphasize the high surface brightness of the arc
however, increasing it by a factor of three.

In short, we see no evidence that component 1 is resolved in the radial
direction.  (In section~\ref{Model} we will argue that the intrinsic
FWHM of the arc is $\sim 0''\!.01$).  Component 5 also appears
unresolved, although its profile suffers from much lower signal to
noise.

Component 2, however, is clearly resolved in figure 2.  To extend the
measurement of component 2's surface brightness profile to larger radii
the image was rotated $180\deg$ about the center of component 2, and
pixels at the locations of other objects in the original image were
replaced with pixels from the rotated image.  This assumes elliptical
symmetry for component 2 in the replaced regions, which cover a maximum
of 25\% (at $r = 1''\!.3$) of the area at any radius, and 7\% of the
total area. Figure 3 shows the resulting radial surface brightness
profile for component 2.  A deVaucouleurs' profile with an effective
radius $r_e \approx1''\!.3$ (10 kpc) provides a much better fit to
component 2 than do exponential disk models, suggesting that this
object is an early type galaxy.  The measured ellipticity of component
2 inside the arc is $\approx 0.16\pm 0.1$ at a position angle of
$\approx 3\pm15\deg$ east of north.  Excess surface brightness appears
near a radius of $1''\!.4$ even though the component 1 pixels (which
are near this radius) have been replaced.  As a check, the surface
brightness profile was measured within sectors centered on component 2
from position angles 73--133$\deg$ and 233--318$\deg$, angles which
bypass all obvious emission sources in figure 1(b).  The value for
$r_e$ in this case was $1''\!.0$ (a smaller $r_e$ is consistent with
these sectors being along the minor axis), and excess light was again
found near $1''\!.4$ radius.  The total excess light at this radius is
very roughly equivalent to a 23rd magnitude source.

\subsection{Photometry} \label{Photometry}

Photometric measurements obtained from the Planetary Camera image for
the components are given in Table 1.  One count in the image
corresponds to $1.185 \times 10^{-21}$ erg/cm$^2$/sec/\AA\, or to a
magnitude of 30.00 in the F814W band with Vega set to magnitude 0.  From
the measured standard deviation per pixel, the sensitivity limit
($3\sigma$) is $m_{814} \sim 28.2$ mag for a point source or $\mu_{814}
\sim 25.6$ mag arcsec$^{-2}$.  Positions are relative to component 2,
whose position in the HST guide star system is given in Table 1.
Polygonal apertures were used to include the faint emission seen
extending from components 1 and 3.  The flux for component 5 was
measured using a $0''\!.35$ diameter aperture, with the local
background measured using the mode of an annulus of width $0''\!.1$
surrounding this aperture, and corrected for PSF losses using the star
H curve of growth.  This flux was checked by subtracting away the image
rotated $180\deg$ about component 2, and also by subtracting the
elliptical model fit to component 2 discussed in
section~\ref{Profiles}.  All three methods consistently gave a value
close to 100 for the flux ratio of component 1 to component 5, and we
adopt 100 for this important ratio for the remainder of the paper.

\section{Discussion}

The morphology of the components of FSC10214+4724, a circular arc
(component~1) with its radius of curvature centered on another object
(component~2), and another fainter image (component~5) on the opposite
side, strongly supports the gravitational lens hypothesis, i.e.  that
component~2 is a foreground galaxy and components~1 and 5 are images of
a single background object.  Under this hypothesis, the multiple
imaging and the arclike morphology and high inferred luminosity of
component~1 result from distortion and magnification by the
gravitational potential of the foreground component~2.  Components~3
and 4 are other galaxies along the line of sight, possibly related to
the galaxy which is component~2, and probably involved in the lensing.

The high resolution of the {\it HST} image makes the arc morphology and
component 5 readily apparent, and allows us to directly measure the
ratio of the brightnesses of these components.  This morphology and
ratio are crucial elements in the development of a lens model for the
source.  We find additional support for the lens hypothesis from the
observed morphology of component~2.  In particular, as shown in
Appendix A, component~2 has the surface brightness profile and spectral
energy distribution expected for a foreground elliptical galaxy, and
its position angle is correctly predicted by the lens model.  In the
following, we adopt the interpretation of FSC10214+4724 as a
gravitationally lensed system, and describe the detailed model of this
system and its consequences.

\subsection{Lens Model} \label{Model}

In the context of a lens model, component~1 is a ``straight arc'' and
component~5 is a ``counterimage.''  This gravitational lens image
configuration is very common; it has been found in several clusters
(see Surdej \& Soucail 1993 for a review).  The model for these systems
is that of a source lying on or very close to a cusp in a caustic (a
line of infinite magnification, e.g.\ Blandford \& Narayan, 1992) in
the source plane.  Although the magnification of a point source lying
on the caustic is formally infinite, the maximum magnification of a
real object is limited by its finite angular radius $r$.  Under the
gravitational lens hypothesis the total magnification of the source
should be on the same order as the flux ratio of arc to counterimage,
roughly $100$ in this case.  Gravitational lens models also predict
that the axis ratio of the arc should be on the same order as the total
magnification.  The $0''\!.7$ length of the arc thus implies an
observed width on the order of $0''\!.007$ (50 pc), or unresolved even
in {\it HST} images.

In the case of lensing dominated by mass at a single
redshift, the gravitational lens mapping, which takes a two-dimensional
angular position $\vec{x}$ on the image plane (i.e.\ the position
observed on the sky) to a two-dimensional angular position $\vec{y}$
on the source plane (i.e.\ the position that would be observed if
there was no lens), is a gradient mapping
\begin{equation}
\vec{y}=\vec{x}-\vec{\nabla}_{\vec{x}}\,\psi(\vec{x}) \;,
\end{equation}
where $\vec{\nabla}_{\vec{x}}$ is the two-dimensional gradient
operator with respect to angular image-plane position $\vec{x}$, and
$\psi(\vec{x})$ is a scaled, projected, two-dimensional gravitational
potential.  The potential is related to the angular surface density
$\Sigma$ (mass per unit solid angle)
\begin{equation}
\Sigma(\vec{x})= \frac{c^2}{8\pi\,G}\,
\frac{D_{\rm d}\,D_{\rm s}}{D_{\rm ds}}\,
\nabla_{\vec{x}}^2\,\psi(\vec{x})\;,
\end{equation}
where $D_{\rm d}$, $D_{\rm s}$ and $D_{\rm ds}$ are angular diameter
distances from observer to lens (deflector), observer to source, and
lens to source, and $\nabla_{\vec{x}}^2$ is the two-dimensional
Laplacian operator.

Where not otherwise stated, the lens models which follow assume that
the potential $\psi$ can be approximated with a quasi-isothermal sphere
with ellipticity (see, e.g.\ Kochanek, 1991), i.e.,
\begin{equation}
\psi(\vec{x})= b\,\sqrt{s^2+r^2}\,
\left[ 1-\gamma\cos 2(\theta-\theta_{\gamma})\right] \:,
\end{equation}
where $\vec{x}=(r,\theta)$ is the position of the point in question
relative to the center of the mass distribution, $b$ is the asymptotic
critical radius (the radius of the Einstein ring), roughly the angular
radius of the circle of images ($\sim 1''$ in this system because that
is the angular separation of arc and lens), $\gamma$ is an ellipticity
parameter, $\theta_{\gamma}$ is the position angle of the major axis,
and $s$ is a core radius.  The results do not depend strongly on the
core radius $s$, so it is assumed to be zero.  The critical radius $b$
can be related to a one-dimensional velocity dispersion for the lens by
\begin{equation}
\sigma_v^2 = \frac{c^2}{4\pi}\,
\frac{D_{\rm s}}{D_{\rm ds}}\, b \;,
\end{equation}
although this depends on the assumption of isothermality.  More secure
is the mass $M$ inside the ``circle of images'' (in this case a circle
of angular radius $b$ around component~2),
\begin{equation}
M= \frac{c^2}{4\,G}\,
\frac{D_{\rm d}\,D_{\rm s}}{D_{\rm ds}}\, b^2 \; .
\end{equation}
The mass $M$ and the inferred luminosity $L$ of the lens can be used to
compute a mass-to-light ratio as well.  The inferred physical
properties of the lens depend strongly on lens and source redshifts and
weakly on world model.  In this system the lens redshift is unknown, so
$\sigma_v$, $M$, and $M/L$ are given in figure~4 as a function of lens
redshift for the model adopted below.  Further discussion of figure~4
is deferred until section~\ref{Redshift}

Model parameters $b$, $\gamma$, and $\theta_{\gamma}$ were varied to
minimize the scatter in the source plane positions corresponding to the
brightest pixels in the arc and counterimage, i.e.:
\begin{equation}
\chi^2 \equiv \sum_i (\Delta \vec{x_i})^2
\end{equation}
where the sum is over the brightest 96 pixels in the deconvolved arc
and the brightest pixel in the counterimage (figure 1(d)), and $\Delta
\vec{x_i}$ is the two-dimensional displacement on the image plane
through which pixel $i$ would need to be moved in order for it to
project (via the lens mapping) to the same location on the source plane
as that of the brightest pixel in the arc.  Image-plane rather than
source-plane displacements were used for computing the scatter because
the image plane is the observed plane, the plane on which uncertainties
are homogeneous and isotropic.  On the source plane the uncertainties
have been mapped through the non-linear lens mapping and are extremely
inhomogeneous and anisotropic.  The minimum rms scatter of the pixels
(in image plane coordinates) was 0.7 PC pixels.

The best-fit model parameters are given in Table~2. The inferred
intrinsic source radius which makes the arc-counterimage flux ratio 100
is $0''\!.0055$ (44 pc).  The model makes the assumption that
component~3 is a singular isothermal sphere ($\gamma=s=0$) at the same
redshift as component~2 and with critical radius $b_3=0''\!.6$, the
expected value under the assumption that components~2 and 3 have the
same mass to ($K$-band) light ratio.  A simpler model which assumes
that the potential is entirely due to an elliptical shaped mass
centered on component~2 was also considered.  The two-component model
was adopted because the intrinsic ellipticity of the potential in this
model is smaller than in the simpler model, in better agreement with
the observed ellipticity in component~2.  This is because the external
mass of component~3 has a tidal effect which replaces some of the
ellipticity in the primary lens.\footnote{If component~3 is at a larger
redshift than component~2 (see the Appendix) the agreement in
ellipticity is slightly better yet, but we adopt a single redshift for
components 2 and 3 to confine the number of parameters.} The predicted
orientation of the lens in the models is consistent with the observed
orientation of component~2.  Figure~1(f) shows the density and
potential contours for the adopted model, as well as the critical curve
and the image morphology for a circular source of radius $0''\!.0055$,
smoothed to a FWHM of $0''\!.02$, and with the counterimage brightness
enhanced for visibility.  Figure ~1(e) shows the model image morphology
convolved with the synthetic PSF discussed in section~\ref{results},
and should be compared to figure~1(c).

The image configuration in the lens model is that of a triple image or
straight arc (plus counterimage).  Although parts of the source are
triply imaged in component~1, the source radius inferred from the flux
ratio of components 1 and 5 is large enough that the three images merge
into a single straight arc. We interpret the peak in the east half of
the arc as corresponding to two images merging on the critical curve,
while the peak in the west half corresponds to the third image.  The
triple structure may become more apparent in high-resolution images in
other bandpasses if the flux at those wavelengths is produced by
structures offset by $\sim 0''\!.02$ (160 pc) from those which produce
the F814W flux, or having intrinsic size scales a factor of $\sim 3$
smaller.

The source location near the point at which three images on one side of
the lens merge into a single image causes high magnification.  As
discussed by Broadhurst and Leh\'ar (1995), the magnification is thus a
sensitive function of source size and position.  The inferred size and
position in turn depend on the assumption of an isothermal profile for
the lens potential, i.e.\ $\psi\propto r$.  For a shallower potential
$\psi\propto r^{0.9}$, the best-fit model puts the center of the source
further inside the three-image region than for the isothermal case, and
the inferred source radius from the arc-counterimage flux ratio is
$0''\!.013$.  For $\psi\propto r^{1.1}$ the inferred source radius is
$0''\!.0046$.

The predicted total magnification of F814W emission from a uniform
circular source as a function of source radius is shown for all three
potential models in figure~5.  In each case, the total magnification
for the source radius derived above from the flux ratio of component~1
to component~5 (i.e. 100) is less or greater than 100 because
component~5 is somewhat demagnified or magnified.  The dependence of
the calculation of the total magnification in the {\it HST} image on
the assumption of a circular source geometry for the F814W emission was
investigated for the isothermal model.  Sources of the same total
projected solid angle on the sky have the same total magnifications to
within about 15\% even if they are highly elliptical, no matter what
their position angle.  The magnification in the isothermal model scales
as $r^{-1}$ for very small sizes or separations from the caustic, and
smoothly converts to $r^{-0.5}$ at larger radii, in agreement with
Schneider, Ehlers \& Falco (1992), and can be approximated to $\sim
20\%$ by ${\bf \rm M} = 3.9 r^{-0.624}$ for the range $0''\!.001 < r <
1"$ (8 to 8000 pc).  The kink at $\sim 0.005$ arcseconds in figure~5
for $\psi\propto r^{0.9}$ occurs where the source size becomes large
enough to make the three distinct images merge into a single arc.
Because in the other two models the source location is closer to the
point at which the three images merge, the source radii at which the
mergers take place are too small to appear in figure~5.  The bump near
$r \sim 0''\!.5$ in figure~5 corresponds to the formation of a ring
(see below).

Different distributions for the narrow line and UV and optical
continuum regions, and the likelihood of substantial reddening (Elston
et al. 1994), can therefore account for the substantially different
appearance of FSC 10214+4724 at different wavelengths noted by Matthews
et al. (1994), and in particular for the larger extent of the $K$-band
arc seen by Matthews et al.  and Graham and Liu (1995) than the arc
seen in the {\it HST} image.  The $140\deg$ extent of the $K$-band arc
corresponds to a source with $0''\!.25$ (2 kpc) radius.   If the source
radius is increased to $\sim 0''\!.5$ it is imaged into an elliptical
($\epsilon \sim 0.4$) ring connecting components~1 and 5. The position
angle of this ring is perpendicular to that of component~2, and is
offset from being perfectly centered on component~2 by $\sim 0''\!.4$
in the direction of component~1.  The excess light near $1''\!.4$
radius noted in section~\ref{Profiles} may be the UV (rest frame)
counterpart of the more extended arc seen in the $K$ images.  Note that
Matthews et al. find the $H\alpha$ emission to be extended in an
east-west direction by $\sim0''\!.5$, suggesting that the narrow line
region is largely coincident with the UV continuum which dominates the
F814W image.

\subsection{Bolometric Luminosity of FSC10214+4724} \label{Luminosity}

FSC10214+4724 has an apparent luminosity of L$_{app}$=
5$\times$10$^{14}$L$_{\sun}$ (Rowan-Robinson et al. 1993), making it
among the most luminous known objects in the Universe.  The vast
majority of this luminosity, $\sim$99\%, is observed in the infrared
(Rowan-Robinson et al. 1991, 1993).  There is strong evidence that the
UV source is a quasar (FWHM of C III] $\sim10,000$ km s$^{-1}$ in
polarized light, Goodrich et al. 1995) enshrouded in dust
(H$\alpha/$H$\beta \ge 20$ implying $A_V > 5.5$, Elston et al. 1994),
and that the quasar's luminosity is absorbed in the dust shell and
reradiated in the infrared (Rowan-Robinson 1993).  This implies that
the size of the infrared emitting region is substantially larger than
the optical/UV emitting region.

If FSC10214+4724 is magnified by a gravitational lens, the intrinsic
source luminosity is less than the apparent luminosity.  However, if
the infrared source is larger than the optical/UV source, the
magnification of the infrared source is less than the magnification
measured from the {\it HST} image.  The magnification of the infrared
source can be estimated by assuming that the infrared source can be
approximated as an optically thick  blackbody.  This assumption
corresponds to making the infrared source as small as possible, and
hence the magnification of the infrared radiation as large as
possible.  In this case, because of the assumption that the emitted
infrared energy distribution is independent of distance from the
central heating source,  the magnification is independent of
wavelength. The temperature of the dust is assumed to be T$\sim$140K.
At this temperature the emission peaks at a rest wavelength of
18$\mu$m, corresponding to the observed emission that peaks at
60$\mu$m.

With this temperature, the apparent luminosity L$_{app}$ and intrinsic
luminosity L$_{int}$ can be written as
\begin{equation}
L_{app} = {\bf
\rm M}(R)L_{int} = {\bf \rm M}(R)\times 4 \pi R^2 \sigma T^4
\end{equation}
where $R$ is the physical radius of the source, ${\bf
\rm M}(R$) is the magnification from figure~5 for a uniform disk of
radius $R$, and $T$ is the blackbody temperature determined by the
wavelength of peak emission.  Solving this equation for $R$ gives a
radius of 130 pc ($0''\!.017$), and ${\bf \rm M}(R) = 42$ for the
isothermal model, so that the intrinisic luminosity of FSC 10214+4724
is $1.2\times 10^{13} L_{\sun}$.   A somewhat larger source size and
lower magnification is derived if the temperature T is assumed to be
115K, the color temperature determined by the observed flux densities
at 60 and 450 $\mu$m and corrected for redshift.  Then the radius of
the infrared source is 240 pc ($0''\!.03$), the magnification is 29,
and the intrinsic luminosity is $1.7\times10^{13} L_{\sun}$.  Note that
at these source radii the magnification is not very sensitive to the
assumed potential (see figure~5).

The expected arc length is $\sim2 r {\bf \rm M}(r)$, or $1''\!.7$ in
the $T=115$K case, and $1''\!.4$ for $T=140$K.  From
VLA-A configuration observations at 8.4 Ghz with $0''\!.25$
resolution, Lawrence et al. (1993) found a $0''\!.6$ (east-west) by
$0''\!.3$ source.  The similarity of this structure to the arc in the
{\it HST} image suggests a continuum radio source radius closer to the
$0''\!.005$ (40 pc) estimated for the optical/UV source than to the
minimum infrared source size just calculated.  Condon et al. (1991)
find that the radio source size for nearby IRAS galaxies with infrared
luminosities $> 10^{12} L_{\sun}$ is typically $\sim 100$ pc (and for
Mrk 231, the most luminous of the sample, $\lesssim 1$ pc), smaller
than the minimum blackbody size for far infrared emission from these
galaxies.  For their sample Condon et al. find $<q> = 2.34$, where $q$
is the logarithm of the ratio of far infrared (60--100$\mu$m) to 1.49
GHz flux.  For FSC10214+4724, extrapolating the Lawrence et al. (1993)
observed radio flux to 0.45 GHz (the observed frequency for emitted
1.49 GHz) yields 3.5mJy, and interpolating to the rest frame
wavelengths for $60$ and $100 \mu$m and using Condon et al.'s
definition gives $q=1.91$.  If the radio magnification is 100, and the
far infrared magnification is 30, then the intrinsic $q = 2.39$.
Therefore the radio morphology and flux measured by Lawrence et al. are
quite consistent with the above estimate for the bolometric
luminosity.

The $0''\!.6$ extent of the radio morphology is also consistent with a
much smaller radio continuum source size, although the value of $q$
would then be significantly larger than observed for local luminous
IRAS galaxies.  It would be interesting (albeit quite challenging) to
see whether the very high angular resolution possible with VLBI
observations revealed the triple structure in the arc discussed above.

The size of the infrared source determined under the assumption of
optically thick emission is a plausible lower limit to the physical
source size. Alternatively, the magnification can be estimated based on
the models of Phinney (1989) of infrared emission from dusty, warped
disks illuminated by a central quasar.  The physical size of the source
required to obtain a self-consistent solution for the intrinsic
luminosity is quite large.  For the region emitting at $150 \mu$m ($450
\mu$m observed), the source radius would be $\sim10''$, much larger
than the observed size of the CO source (e.g. Scoville et al.  1995).
Thus we consider such a model less consistent with the observations
than the optically thick models.

The reduction in the intrinsic luminosity of FSC10214+4724 to $\sim 2
\times 10^{13} L_{\sun}$ implied by the lens model of the source brings
it into the luminosity range of previously studied infrared luminous
AGN.  The IRAS source FSC15307+3252 at a redshift of z=0.93 has a
luminosity of $4\times10^{13} L_{\sun}$, while the IRAS source
PSC09104+4109 has a luminosity of $2\times10^{13} L_{\sun}$ for our
assumed cosmology (Cutri et al. 1994).  There is no known evidence from
high resolution imaging (Soifer et al. 1994, Hutching and Neff 1988,
Soifer et al. 1995 in preparation) that either of these sources is a
gravitational lens, so the apparent luminosity is presumably the
intrinsic luminosity in these cases.  Thus, based on its bolometric
luminosity,  FSC10214+4724 is most likely a source simlar to these.
The reduction in intrinsic luminosity reduces the necessary dust mass
associated with the source (Rowan-Robinson et al. 1993) by the same
magnification factor, into the range $M_{dust} \sim 1-3 \times 10^7
M_{\sun}$, which is consistent with the estimates of the gas mass based
on the dynamical mass determinations from the CO observations (Scoville
et al. 1995).

\subsection{Properties of Component 2} \label{Redshift}

No conclusive measurement of the redshift for component~2 has yet been
made, to our knowledge, although tentative values of 0.42 (Close et al.
1995) and 0.90 (Serjeant et al. 1995) have been suggested based on
possible continuum breaks in the spectrum of component~2, while
Goodrich et al. (1995) find Mg lines in absorption at $z=1.32$ (and
possibly $z=0.89$) in the spectrum of component~1.  In the Appendix we
provide three estimates of the redshift for component~2 (two of which
are closely related).  All three estimates are consistent with $z \sim
0.9$, and we adopt this value as as our best overall estimate of the
redshift.  Note the SED and $R_e - <\mu_B>_e$ estimates do not assume
component~2 is a lens, only that it is an elliptical galaxy, and
therefore give additional support to the lensing hypothesis by placing
component~2 at an intervening redshift relative to FSC10214+4724.

The velocity dispersion $\sigma_v$, mass $M$, and mass-to-light ratio
$(M/L)$ predicted for the lens are shown in figure~4 as a function of
lens redshift.  Adopting $z=0.9$ yields $(M/L)_B = 8 M_{\sun}/L_{\sun}$
(vs. the observed average of $6 M_{\sun}/L_{\sun}$, van der Marel
1991), $\sigma_v = 270$ km s$^{-1}$, and $M = 3.9 \times 10^{11}
M_{\sun}$ (thus $L_B = 5\times10^{10} L_{\sun}$).  These values are for
a radius of $0''\!.85$: using figure~3 the total blue luminosity is
then $L_B = 1.4\times10^{11} L_{\sun}$ or $\sim 4 L*$ (Binggeli,
Sandage and Tammann 1988).  These values are independent of
evolutionary model because F814W samples rest frame $B$ at $z = 0.9$.
The velocity dispersion and mass estimated by Graham and Liu (1995),
Broadhurst and Leh\'ar (1995), and Close et al. (1995) are consistent
with figure~4, but their total luminosity is lower (and hence $(M/L)_B$
higher) because a smaller aperture correction than is shown in figure~3
was assumed.

Thus for the redshift estimate $z=0.9$ the present lensing model
predicts properties typical of present day elliptical galaxies, except
that the galaxy is unusually luminous.  The probability of a large
lensing galaxy is greater than the galaxy luminosity function alone
implies, however, because the crossection for gravitational lensing is
proportional to mass.

\subsection{The Parent Population of IRAS FSC10214+4724}

Analysis of statistically complete samples of radio galaxies suggests
that the lensing rate (i.e. probability that a given radio galaxy is
lensed) is on the order of $1/500$ (Miralda-Escud\'e \& Leh\'ar 1992,
Myers et al. 1995).  Given that a source is lensed, the probability of
getting total magnification ${\bf \rm M}$ is on the order of ${\bf \rm
M}^{-2}$ (e.g.\ Schneider, Ehlers \& Falco, 1992).  The estimated total
magnification $\sim 30$ for the IRAS flux from section~\ref{Luminosity}
corresponds to a likelihood of $\sim 10^{-3}$.  The existence of a
single lensed object in the surveyed area ($0.2$~sr - Rowan-Robinson
1991) with magnification 30 should, according to these probabilities,
represent an underlying population of $\sim800$ compact, $60
\mu$m-luminous objects per square degree (or $>40$ per square degree at
$95\%$ confidence) which are either not lensed or lensed with much
lower magnification (and hence are not in the FSS catalog).  If they
are like IRAS FSC10214+4724, these sources will have observed magnitude
$r\sim 25$ mag, and their IR fluxes will be of order 3mJy at $25 \mu$m
and 7 mJy at $60 \mu$m.  To these flux levels, models of the IR galaxy
population with strong luminosity evolution (Hacking and Soifer, 1991)
predict a few hundred sources per square degree, in agreement with this
estimate.  Of course this is only an order of magnitude estimate
because it depends on extrapolation from a single serendipitously
discovered object, and on the relative redshift distributions of
IR-luminous and radio galaxies.  Optical field galaxy redshift surveys
now underway with the Keck Telescope are approaching this depth (J.
Cohen, private communication; UC DEEP collaboration, private
communication), and IR imaging surveys to well beyond these levels are
envisioned with ISO, WIRE, and SIRTF, so this very uncertain prediction
may be testable in the near future.

\section{Summary}

We have obtained a $0.8\mu$m image of the $z=2.286$ IRAS source
FSC10214+4724 with the {\it HST} WFPC2 Planetary Camera, with $0''\!.1$
resolution and high signal to noise.  We find the following:

1) The source appears as an unresolved ($< 0''\!.06$ wide) arc
$0''\!.7$ long, with significant substructure along its length.  The
arc is roughly centered on a galaxy $1''\!.18$ to the north
(component~2), and a faint unresolved component (component~5) is
clearly detected $0''\!.43$ north of component~2.  Two other galaxies
(components 3 and 4) are evident within a few arcseconds of the IRAS
source.  This morphological configuration is characteristic of a
gravitationally lensed system, in which the arc and component~5 are
images of a single background source produced by the potential of the
foreground component~2.

2) The surface brightness profile of component~2 is well matched by a
de Vaucouleurs profile, characteristic of an elliptical galaxy with an
effective radius of $1''\!.27$.  There is evidence for excess emission
above the de Vaucouleurs profile near the radius of the arc.

3) The flux ratio of the arc to the component~5 is $\sim 100$, implying
magnification in the {\it HST} image of the background source by
roughly this amount.

4) A detailed lensing model, which reproduces the observed morphology
and relative flux of the arc and counterimage, correctly predicts the
position angle for component~2.  Better agreement is found with the
observed ellipticity of component~2 if component~3 is included in the
lensing potential.  The model predicts reasonable values for the mass
and velocity dispersion of component~2.

5) If component~2 is an elliptical galaxy, its spectral energy
distribution is inconsistent with it being at $z=2.286$, and $z=0.9$ is
preferred.  The surface brightness profile of component~2 implies a
redshift between 0.6 and 1.2.  From the lensing model, for $z \sim
0.9$, the central mass-to-light ratio for component~2 is $(M/L)_B = 8
M_{\sun}/L_{\sun}$, the velocity dispersion $\sigma_v = 270$ km
s$^{-1}$, and the total blue luminosity $L_B = 1.4 \times 10^{11}
L_{\sun} \sim 4 L*$.

6) The model predicts an intrinsic radius of $\sim 0''\!.005$ (40 pc)
for the background source at $0.25 \mu$m rest wavelength.  Triple
structure in the arc is obscured by this source size, but may become
apparent at high resolution in other bandpasses.  The larger size of
the arc observed at $K$ implies an intrinsic source radius of
$0''\!.25$ in the corresponding emitting bandpass.  A source of radius
$> 0''\!.5$ would produce a ring of emission connecting the arc and
component~5.  This may account for the excess emission seen in the
surface brightness profile of component~2.  The $H\alpha$ and radio
continuum morphologies appear similar to that of the $0.8\mu$m arc,
implying a similar source size for the narrow line, UV continuum, and
radio continuum emission.

7) The minimum source size for an optically thick blackbody source
producing the bulk of the bolometric luminosity is $\sim0''\!.03$ (240
pc), implying  a bolometric magnification of $\sim30$.   The background
lensed source then has an intrinsic luminosity $\sim 2 \times 10^{13}
L_{\sun}$.  Thus IRAS FSC10214+4724 is not the most luminous object in
the Universe, but it remains among the most luminous in the IRAS
catalog.

8)  The expected incidence of 30-fold gravitational magnification is
low enough to suggest that FSC10214+4724 represents an underlying
population of $\sim800$ compact objects per square degree with optical
magnitude $r\sim 25$ mag and $F_{60\mu{\rm m}} \sim 7$mJy.

\clearpage

\acknowledgments

We thank Mark Dickinson for help with the $R_e - <\mu_B>_e$ technique
for estimating $z$ and in particular for supplying the Sandage \&
Perelmutter data in electronic form, Adam Stanford for calculating
K-corrections and general assistance with STSDAS, and Roger Blandford
for help with lens modelling.  We acknowledge helpful discussions with
James Graham, Michael Liu, Tom Broadhurst, Joseph Leh\'ar, and Joseph
Miller.  The ideas of {\it rotating} (rather than merely flipping)
component 2 in section~\ref{Profiles} and of a magnification vs. radius
plot (cf figure 5) were suggested by Broadhurst and Leh\'ar.  An
anonymous referee reminded us of the sensitivity of the derived
magnification to the isothermal profile assumption.  This research was
supported by NASA through a grant awarded by STScI, which is operated
by AURA under NASA contract NAS 5-26555.  Portions of the research
described in this paper were carried out by the Jet Propulsion
Laboratory, California Institute of Technology, under a contract with
NASA.

\clearpage

\appendix

\section{Estimates of the Redshift for Component~2}

Here we use the spectral energy distribution and surface brightness
profile of component~2 to estimate its redshift.

\subsection{Spectral Energy Distribution} \label{SED}

Figure 6 combines $R$ and $H$ data from Elston et al. (1994), $J$ and
$K$ from Matthews et al. (1994), and F814W data from the present work
for components 2, 3, and 4, normalized at $K$.  (Note that component~1
in the Elston et al. terminology is our component~2.) Because the
angular resolution of the three data sets ranges from
$0''\!.1$--$1''\!.5$, the combined spectral energy distribution (SED)
is somewhat uncertain.

Given the good agreement of the surface brightness profile for
component~2 with a de Vaucouleurs law (figure 3) it is reasonable to
assume that this component is an elliptical galaxy. For comparison, the
unevolved spectrum of a standard Bruzual and Charlot (1993) elliptical
galaxy model at an age of 13 Gyr and redshifts of 0.42, 0.90, 1.32, and
2.286 is plotted in figure 6(a); the corresponding passively evolving
model with ages of 7.75, 5, 3.75, and 2.2 Gyr at these redshifts (the
ages are consistent with a present age of 13 Gyr with the assumed
cosmology) is plotted in figure 6(b).  All models were normalized to
the $K$ flux in the SED for component~2.  Clearly the $z=2.286$ models
fail to match the observed SED for component~2, while the $z=0.9$
models provide surprisingly good fits to the observations.  This is
fairly strong evidence that component~2 is in fact a foreground
elliptical: it is too blue to be an elliptical galaxy at the redshift
of FSC10214+4724.

The $r$ data point for component 3 is anomalously bright, while the
rest of its SED is somewhat redder than component~2.  This might be due
to a combination of reddening and star formation associated with the
tidal interaction suggested in section~\ref{Morphology}, placing
component~3 at the same redshift as component~2, as the adopted model
in section~\ref{Model} assumes.  Alternatively, the $z=1.32$ SED models
appear to fit component~3 at least as well as the $z=0.9$ models.
Without the constraint of an elliptical surface brightness profile, it
is much more difficult to assign a unique redshift based on the SED for
component~3 than for component~2.  Putting component~3 at $z=1.32$
implies a higher lensing mass, for a constant mass-to-light ratio, and
the result is a decrease in the mass and ellipticity of component~2 by
of order 10\%.  Component~4 is significantly bluer than components~2
and 3, consistent with what appears to be a later type morphology.

\subsection{Fundamental Plane Relations} \label{Dnsigma}

Since the surface brightness profile of component 2 strongly suggests
it is an elliptical galaxy (section~\ref{Profiles} and figure 3), it is
possible to make further use of the surface brightness profile to
estimate the redshift of component~2 using the fundamental plane
relations for ellipticals (Kormendy and Djorgovski 1989).  Using
$r_e=1''\!.274$ and $m_{814}=21.04$ within $r_e$ from figure 3, the
present day equivalent blue surface brightness of component 2 was
calculated by correcting for $(1+z)^4$ surface brightness dimming,
redshift K-correction, and luminosity evolution.  Figure 7 shows $R_e$
(in kpc) and $<\mu_B>_e$ (the average blue surface brightness within
$R_e$) as a function of the assumed redshift for component 2, overlaid
on the data for present day ellipticals from Sandage and Perelmutter
(1990).  Luminosity and K-corrections are shown for both a non-evolving
and passively evolving elliptical model spectrum from Bruzual and
Charlot (1993).  Redshifts near zero, or in the range 0.6 to 1.2 can be
accomodated.  From figure 4, the lens model predicts a central
mass-to-light ratio $> 100$ for $z < 0.2$, arguing against low values.
We consider the passively evolving model more realistic, leading to an
estimate of $z=1.0\pm0.2$.  This estimate is independent of $H_o$
because the present day data scale in the same way as the calculated
values.  The estimate is driven almost entirely by the $(1+z)^4$
dependence of surface brightness on redshift, and is relatively
independent of $q_o$ because the latter primarily affects angular size,
which is nearly orthogonal to redshift in the region of interest in
figure 7.  The main uncertainty is due to luminosity evolution (some of
which arises from the dependence of timescales on $q_o$) and to the
scatter in surface brightness among giant elliptical galaxies.

Much of this scatter is correlated with the velocity dispersion, and if
$\sigma$ is known the $D_n - \sigma$ relation (Lynden-Bell et al. 1988)
can be used to measure the angular diameter distance for giant
elliptical galaxies.  Here $D_n$ is defined as the angular diameter of
the circle within which the integrated rest frame blue surface
brightness is 20.75 mag arcsec$^{-2}$ after correction for luminosity
evolution and $(1+z)^4$ surface brightness dimming.  The value of
$\sigma$ is found from the lens model as plotted in figure 4(a).  An
advantage of $D_n$ over $R_e$ is that $D_n$ is defined at a higher
surface brightness level, and is therefore smaller than $R_e$ and more
immune to uncertainties about emission at the arc radius, as well as
being less sensitive to uncertainties in sky subtraction.  However the
technique is sensitive to $q_o$ because it is essentially an angular
diameter distance.  The redshift estimates from this approach range
from 0.75 for the case $q_o=0$ and no evolution, to 1.15 for $q_o=0.5$
and passive evolution, independent of $H_o$.

\clearpage

\begin{table*}
\begin{center}
\begin{tabular}{lrrrl}
Component  & $m_{814}$ & $\Delta\alpha ('')$ & $\Delta\delta ('')$ &
Comment\\
\tableline

1                &   20.44 & 0.10 & -1.16 & includes faint extensions\\
2   &  21.41 & 0 & 0 & inside $r=0''\!.85$ \\[-1ex]
2           & 20.3 & & & total\\
3     & 23.16 & 1.03 & 1.93 & inside $r=0''\!.5$ \\[-1ex]
3      & 22.98 & & & including component to east  \\
4                 &  23.58 & 3.42 & 1.79
& $\sim1''\!.1$ wide $\times 0''\!.6$ high polygonal aperture\\
5                  & 25.5 & 0.03 & 0.43 & inside $r=0''\!.35$\\
Star H             & 24.54 & -6.59 & -6.20 \\
\end{tabular}
\end{center}
\caption{Photometry of objects in HST F814W image of
IRAS FSC 10214+4724.}
\tablecomments{
Components are identified in figure 1(a) and (b).  Positions are with
respect to the center of component 2, which is approximately
$\alpha=10{\rm h}24{\rm m}34^{\rm s}\!.56,
\delta= 47\deg09'10''\!.8,$ (J2000)
in the {\it HST} guide star catalog frame.
For component 1 the position is for the peak brightness.}
\end{table*}

\begin{table*}
\begin{center}
\begin{tabular}{cccl}
Parameter                  & Model Value    & Observed & Comments \\
\tableline
$b$                         & $0.82$    & & arcseconds \\
$\gamma$                    & $0.12$    & & defined by Eqn.~(3)\\
$\epsilon$                  & $0.30$    & $0.16 \pm 0.1$ & $1-{\rm b/a}$ \\
$\theta_{\gamma}$           & $-11$     & $3 \pm 15$ & degrees, N of E \\
$b_3$                       & $0.60$    & & arcseconds \\
source radius               & $0.0055$  & &
arcseconds, for magnification ratio 100\\
\\

\end{tabular}

\end{center}
\caption{Isothermal Lens Model Parameters.}
\tablecomments{The model assumes the lensing potential arises from
components 2 and 3, and that these components have isothermal
potentials and the same redshift and mass to ($K$-band) light ratios.
The ellipticity $\epsilon$ is the conventional value defined by one
minus the ratio of semiminor to semimajor axis, and differs from the
model ellipticity parameter $\gamma$ which is defined by Eqn.~(3).  The
``source radius'' is the angular radius at which, for a circular
source, the arc-counterimage magnification ratio is 100.  Other symbols
are explained in the text.  World model $q_0=0.5$ is assumed.  Changing
world models only changes the numbers by $\sim 10\%$.}

\end{table*}

\clearpage

\noindent

\begin{figure}
\caption{ Montage of HST Planetary Camera (PC) imaging of IRAS
FSC10214+4724 in F814W. Panels (a), (b), and (c) show the image at
progressively finer scales, as indicated by the axes which are labelled
in arcseconds relative to component 2 (see panel (b)).  Panels (d),
(e), and (f) have the same scale and center as panel (c).  Panel (d) is
a partially deconvolved version of the data with a factor of four
subsampling.  Panel~(e) shows the observed image configuration given
the lens model and a uniform circular disk source of radius
$0''\!.0055$.  The image configuration has been convolved with the
synthetic {\it HST} PSF and binned into PC pixels to allow direct
comparison with panel~(c).  Panel (f) shows further details of the lens
model:  as in panel (d) the pixel size is four times smaller, and the
model image has been lightly smoothed to a FWHM of $0''\!.02$.  Lines
indicate contours of mass (dotted), potential (dashed), and the
critical curve (solid line) for the model.  The grey levels in panel
(f) range linearly from zero (white) to the peak value in the arc
(black), but have been enhanced (in panel (f) only) by a factor of
seven at the counterimage location.  The grey levels in panels (a) --
(e) range linearly from 0.5\% to 5\% of the peak brightness in
component 1 ($\mu_{814}=17.6$ mag arcsec$^{-2}$ in panels (a) -- (c),
$\mu_{814}=13.6$ mag arcsec$^{-2}$ in panel (d)).  Contour levels in
panels (c) -- (e) are at 25, 50, 75 and 90\% of this peak brightness
for component 1, or at 25, 50, 75 and 90\% of the peak brightness for
component 2 ($\mu_{814}=19.4$ mag arcsec$^{-2}$ in panel (c),
$\mu_{814}=15.6$ mag arcsec$^{-2}$ in panel (d)), as appropriate.
North is $37.1\deg$ counterclockwise from vertical in all panels, with
east $90\deg$ counterclockwise from north, as shown in panels (a) and
(b).}
\end{figure}

\begin{figure}
\plotfiddle{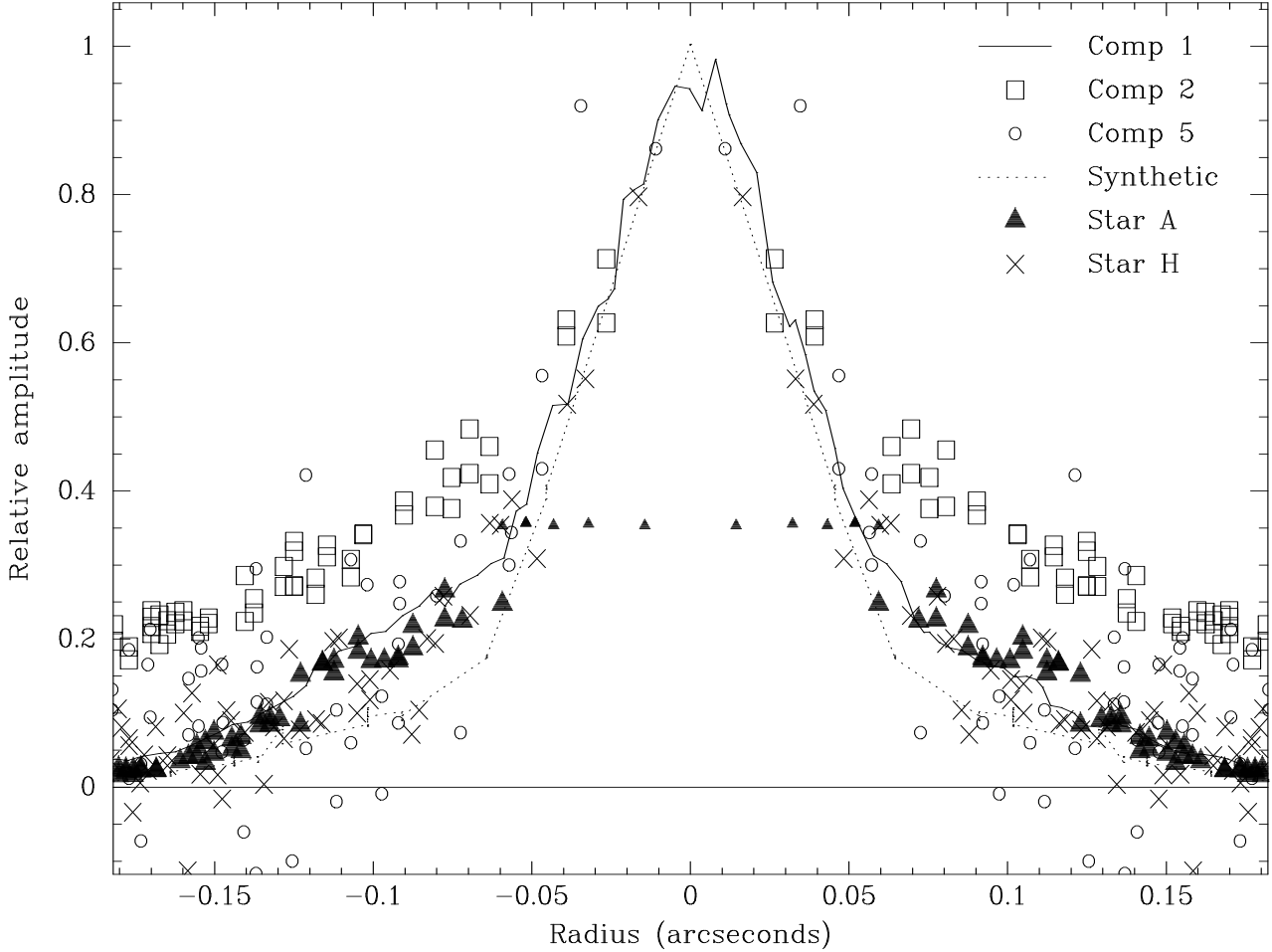}{4in}{270}{70}{70}{-288}{432}
\caption{Radial profiles for objects identified in figures 1(a) and
(b).  Component~1 appears unresolved relative to stars A and H.  For
component~1 the equivalent radial profile is plotted, as discussed in
section 3.2. The synthetic PSF was used to generate the deconvolution
shown in Figure 1(d).  The vertical scales for the profiles were
normalized at the smallest radius available, except for star A, whose
core is saturated in our image, and which was normalized to the
synthetic PSF at the first radius which was not saturated.  The data
points plotted for negative radii are identical to those for positive
radii except for component 1.}

\end{figure}

\begin{figure}
\plotfiddle{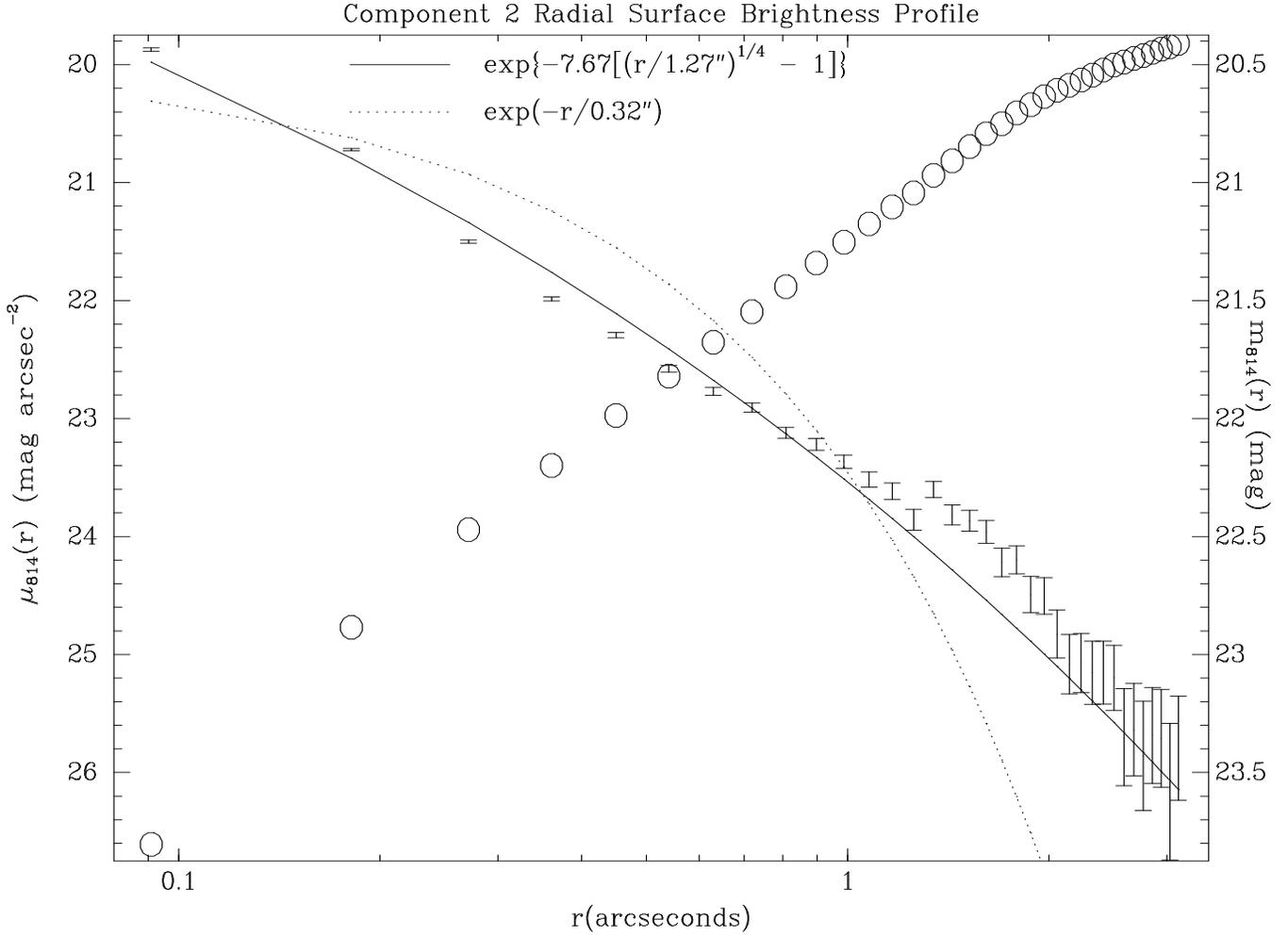}{4in}{270}{70}{70}{-288}{432}
\caption{The surface brightness of component 2 is plotted as a function
of radius, together with a deVaucouleurs profile with $r_e = 1''\!.27$
(solid line) and an exponential profile with $r_d = 0''\!.32$ (dotted
line).  Magnitude as a function of aperture radius is shown by the open
circles and the righthand scale.  The magnitude appears to converge
near $m_{814} = 20.3$ mag, yielding a consistent value for $r_e$.}
\end{figure}

\begin{figure}
\plotfiddle{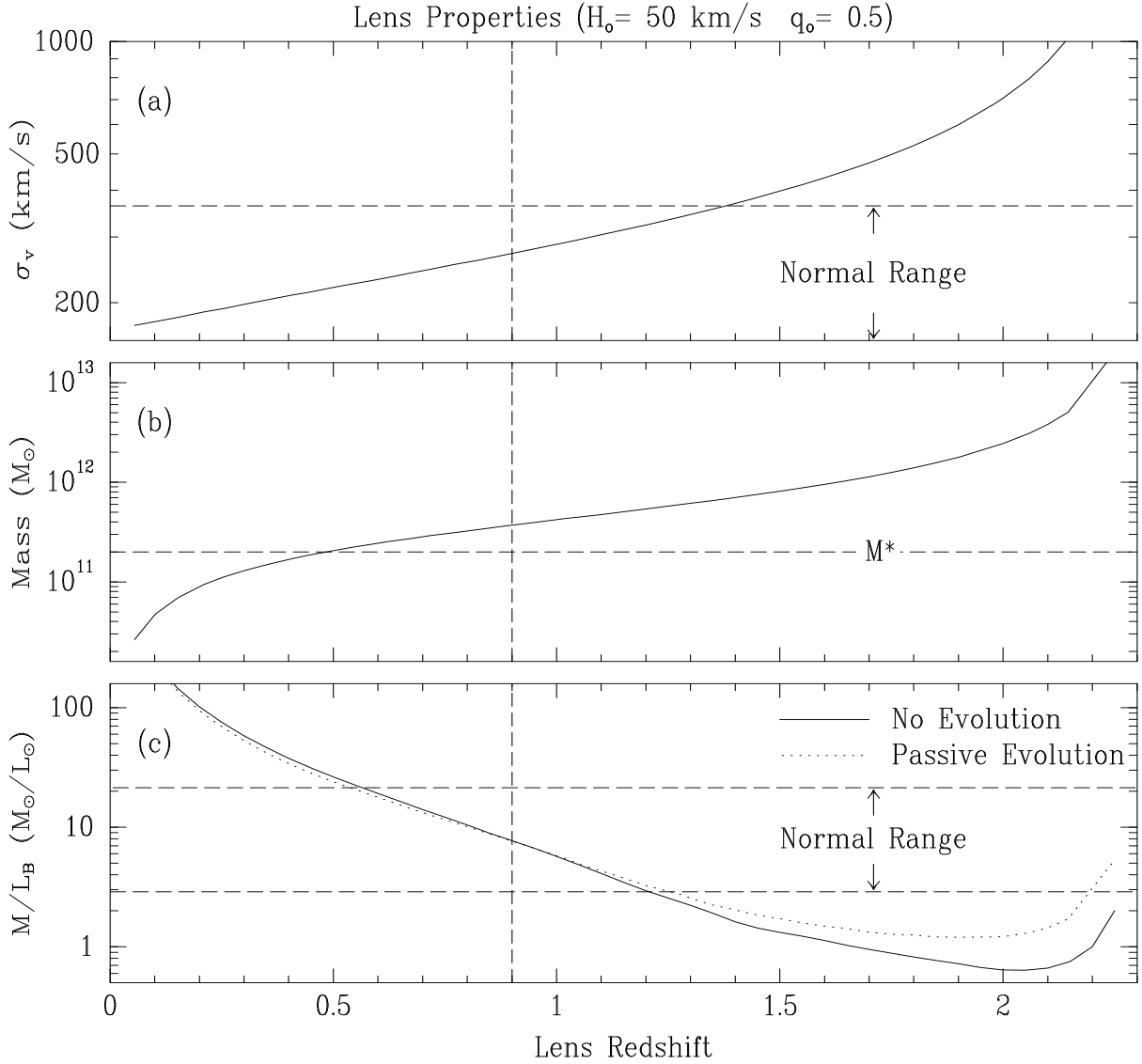}{6in}{270}{70}{80}{-288}{470}
\caption{Predictions from the lens model (see section 4.1 and Table 2)
for the mass inside the critical radius (panel b) and the velocity
dispersion (panel a) of the lens as a function of its redshift.  The
bottom panel (c) shows the mass to luminosity ratio using the {\it HST}
F814W flux inside $0''\!.85$ radius, K-corrected to rest frame $B$
using unevolving (solid curve) and passively evolving (dotted curve)
elliptical model spectra from Bruzual \& Charlot (1993).  The dashed
vertical line indicates the best estimate for the redshift of
component~2 as discussed in the Appendix.  Horizontal dashed lines show
the normally observed range of $(M/L)_B$ and $\sigma_v$ from Fisher,
Illingworth, and Franx (1995), van der Marel (1991), and Davies et al.
(1983).  The value $M^* = 2\times 10^{11} M_{\sun}$ is from van der
Marel's mean $(M/L)_B = 6 M_{\sun}/L_{\sun}$ and $L^*_B = -21$ mag from
Binggeli, Sandage and Tammann (1988).  The values shown are for the
assumed cosmology ($H_o = 50$ km s$^{-1}$, $q_o = 0.5$), and are
relatively insensitive to $q_o$. The mass scales as $h^{-1}$ and the
mass to luminosity ratio as $h$ (for the unevolving K-correction),
while the velocity dispersion is independent of $h$, where $h$ is the
Hubble constant in units of 100 km s$^{-1}$ Mpc$^{-1}$.}
\end{figure}

\begin{figure}
\plotfiddle{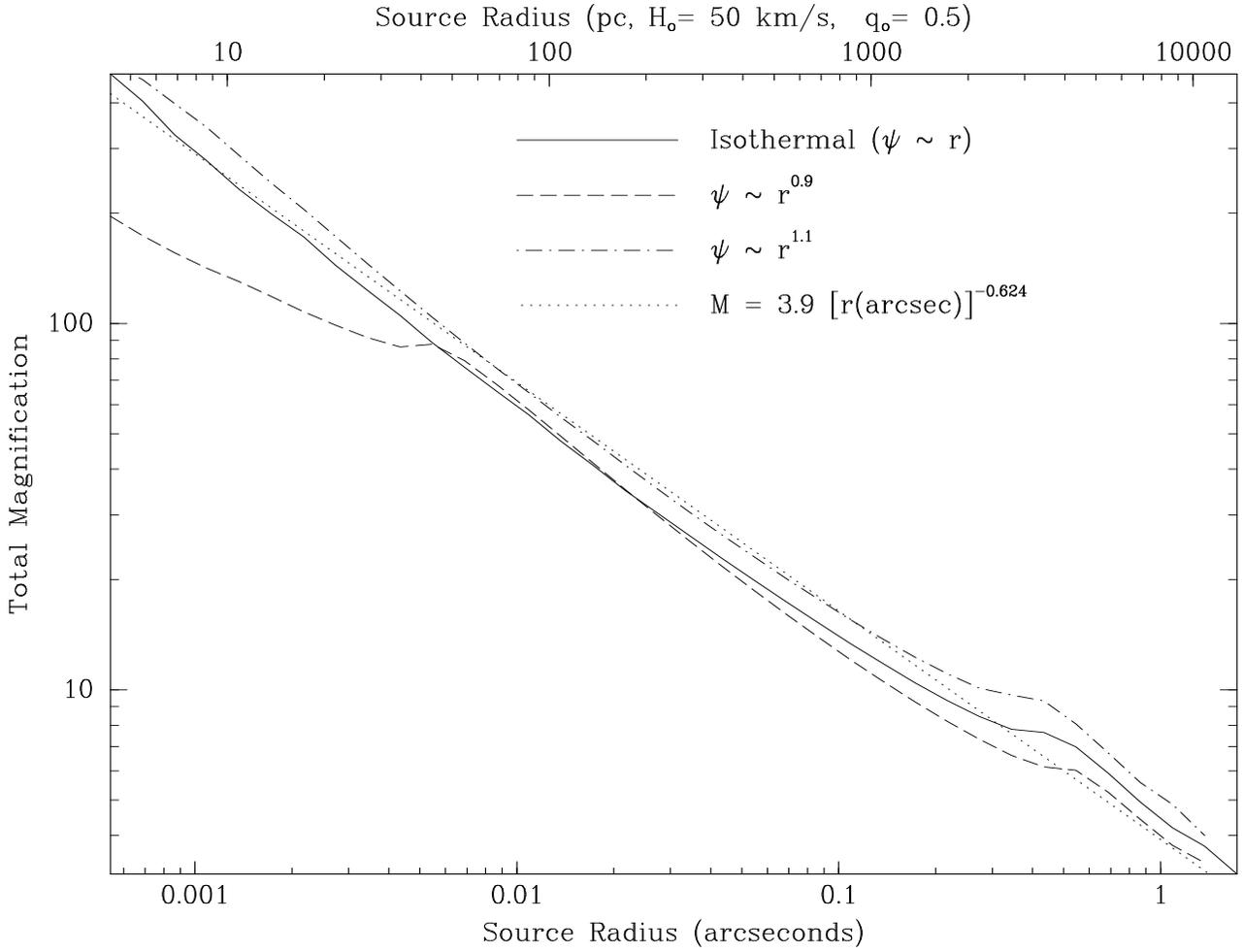}{4in}{270}{70}{70}{-288}{432}
\caption{Predictions from the lens model for the total magnification
${\bf \rm M}$ of the background source flux, assuming a uniformly
illuminated source of radius $r$ (arcseconds, shown on the bottom axis)
or $R$ (parsecs, shown on the top axis).  Different line types
correspond to different assumed potentials as shown in the legend.  The
dotted line shows a power law approximation to the predicted
magnification.}
\end{figure}

\begin{figure}
\plotfiddle{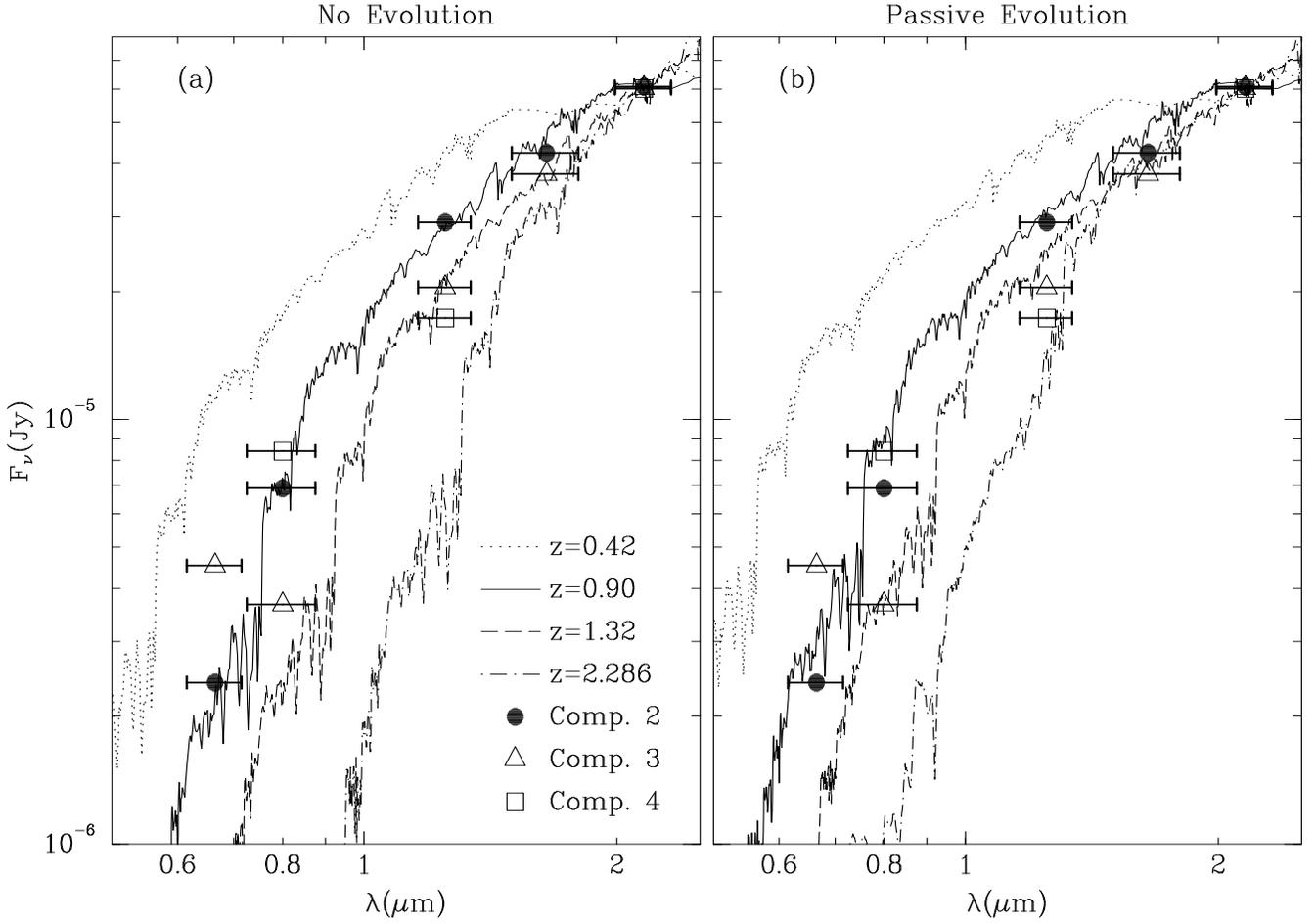}{4in}{270}{70}{70}{-288}{432}
\caption{Spectral energy distributions for components identified in
figure 1(b), with unevolving (panel a) and passively evolving (panel b)
model elliptical spectra from Bruzual \& Charlot (1993) shown for
comparison.  The component~2 data fit the $z=0.9$ models well. The flux
scale is correct for component~2; for component~3 it should be reduced
by a factor of 2.3 and for component~4 by a factor of 9.  The data are
derived from Elston et al. (1994), Matthews et al. (1994), and this
paper.  See section A.1 for further details.}
\end{figure}

\begin{figure}
\plotfiddle{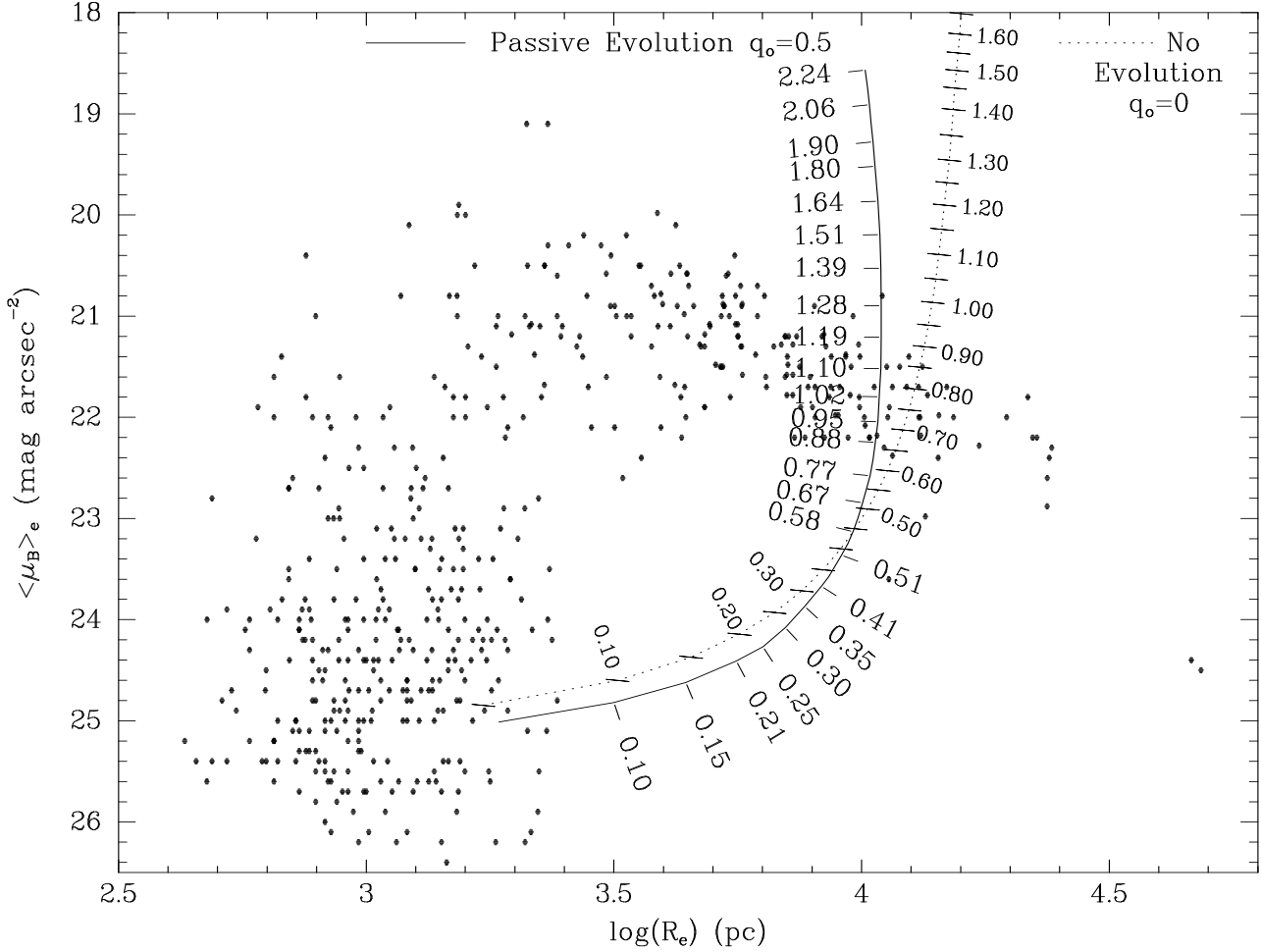}{4in}{270}{70}{70}{-288}{432}
\caption{Estimate of the redshift for component 2 using the average
blue surface brightness - effective radius relation for giant
elliptical galaxies.  The data are from Sandage and Perelmutter
(1990).  The dotted curve shows calculated $<\mu_B>_e$ and $R_e$ values
with K-corrections from an unevolving Bruzual and Charlot (1993)
elliptical model with $q_o =0$ and a uniform redshift interval of
0.05.  Here $<\mu_B>_e$ is the average blue surface brightness within
the effective radius.  The solid curve is for a passively evolving
elliptical and $q_o = 0.5$.}
\end{figure}

\end{document}